# Thermal expansion in 2D honeycomb structures: Role of transverse phonon modes

Sarita Mann and V.K. Jindal[1]

Department of Physics, Panjab University Chandigarh-160014.

Graphene and its derivatives including hexagonal BN are notorious for their large negative thermal expansion over a wide range of temperature which is quite unusual. We attempt to analyze this unusual behavior on the basis of character of the phonon modes. The linear thermal expansion coefficients (LTEC) of two-dimensional honeycomb structured pure graphene, *h*-BN and B/N doped graphene are studied using density functional perturbation theory (DFPT) under quasi harmonic approximation. The dynamical matrix and the phonon frequencies were calculated using VASP code in interface with phonopy code. The approach is first applied to pure graphene to calculate thermal expansion. The results agree with earlier calculations using similar approach. Thereafter we have studied the effect of B/N doping on LTEC and also compared it with LTEC of h-BN sheet. The LTEC of graphene is negative in the whole temperature range under study (0-1000K) with a room temperature (RT) value of $-3.51 \times 10^{-6} K^{-1}$. The value of LTEC at room temperature becomes more negative with B/N doping in graphene as well as for h-BN sheet. In order to get an insight into the cause of negative thermal expansion, we have computed the contribution of individual phonon modes of vibration. We notice that it is principally the ZA (transverse acoustic) mode which is responsible for negative thermal expansion. It has been concluded that transverse mode in 2D hexagonal lattices have an important role to play in many of the thermo dynamical properties of 2D structures.

**Keywords**- Graphene, h-BN, density functional perturbation theory, phonons, doping, thermal expansion, Grüneisen parameter

## 1. INTRODUCTION

The first experimental synthesis of graphene [1], a one atom thick thermodynamically stable structure of carbon, led to a lot of research interest. In graphene $sp^2$ hybridized carbon atoms are arranged in honeycomb lattice structure. It has attracted great scientific interest mainly because of its unusual electronic properties. Graphene is a very promising candidate for several future applications because of its ballistic transport at room temperature combined with chemical and mechanical stability [2]. Its charge carriers exhibit very high intrinsic mobility, electron transport is governed by a Dirac-like equation, so it acts like a bridge between the condensed matter physics and electrodynamics [3-5]. There has been a lot of research ongoing to study and modify electronic and optical properties of graphene by doping with a variety of atoms. We in our previous research [6-8] have studied the band gap modification and optical behavior by substituting the carbon atoms with boron and nitrogen atoms.

---

[1] Present address – Institute of Physical and Theoretical Chemistry, Julius-Maximilians University Würzburg, Germany. Email for correspondence: Jindal@pu.ac.in or vijay.jindal@uni-wuerzburg.de

Equally important aspect of graphene is its unique thermal properties. It has unusually high thermal conductivity [9] which is a subject of intensive research. Heat flow hindrance in miniature components is a matter of concern in the micro and nano-electronics. There is a lot of research ongoing in this area. In addition to this, graphene also has unusual negative linear thermal expansion coefficient (LTEC) as reported by several authors [10-15]. Kim et al [16] is another important study which focuses on negative thermal expansion of graphyne. Wejrzanowski *et al.* [17] report a study on copper-graphene composites and find interesting anisotropy in thermal conductivity in planner direction and transverse direction. Mounet *et al.* [10] using first-principles based quantum espresso code under quasi harmonic approximation (QHA) reported that LTEC of graphene is negative upto temperatures as high as 2500K with room temperature value around $-3.6 \times 10^{-6}$ K$^{-1}$. Sevik [11] has done the calculations using DFPT implemented in VASP predicting almost same behavior as Mounet *et al.* On the other hand experimentally, Yoon *et al.* [12] have estimated LTEC of single layer graphene to be negative in whole temperature range under study i.e. 200-400K using temperature-dependent Raman spectroscopy, with a room temperature value of $-8 \times 10^{-6}$ K$^{-1}$. However, Zakharchenko *et al.* [13] estimated the average value to be $-4.8 \times 10^{-6}$ K$^{-1}$ between the temperatures ranges 0–300 K and predicted a negative trend up to only 900 K by Monte Carlo simulations, based on the empirical bond order potential. In another experimental study, Bao *et al.* [14] determined the LTEC of graphene as negative only up to 350 K and measured the room temperature value as approximately two times larger than the theoretical estimations, $-7 \times 10^{-6}$ K$^{-1}$. In addition, Jiang *et al.* [15] determined the LTEC of graphene using non-equilibrium green's function with and without substrate with a large negative region at low temperatures and very small value at high-temperature limit, and the value of LTEC at 300 K is $-6 \times 10^{-6}$ K$^{-1}$ which is very close to experimental result.

Further, it has been brought to our notice a comprehensive work on the anharmonicity of graphene exploring in detail the thermal expansion of graphene. These works are fundamentally interesting but do not focus on the salient feature of transverse modes in comparison to acoustic modes causing negative thermal expansion [18-20].

The h-BN (hexagonal-boron nitride) sheet is another two-dimensional (2D) material which has attracted great attention in recent years after graphene because of its excellent properties and applications [21, 22]. Crystal lattice of h-BN consists of a sp$^2$ bonded 2D hexagonal lattice with boron and nitrogen atoms occupying alternate positions similar to graphene. Boron and nitrogen atoms are bound by strong covalent bonds within the h-BN sheet. The elastic constants are very similar, but the polar nature of the BN bond leads to significant changes in the electronic structure of h-BN as compared to graphene. While graphene is a semimetal (zero band gap semiconductor), h-BN has a large band gap (5-6 eV) [23]. Further its high thermal stability, high dielectric breakdown and relative chemical inertness distinguishes it from graphene. These properties make it a good choice for many industrial applications. Because of the high chemical and thermal stability of h-BN, its ceramics are extensively used as a part of high temperature devices and also in gas seals of oxygen sensors, crucibles for molten glasses and melts and parts of high temperature furnaces also etc. Therefore, it follows naturally that the unique electronic properties are supplemented by a study of their thermodynamic and phonon related properties as well, only then the full potential in the applications of graphene or doped graphene or BN sheet is exploited. Sevik [11] has also studied LTEC of h-BN sheet using VASP software and reported negative thermal expansion with almost double the room temperature value as compared to graphene. Thomas *et al.* [24] have studied the LTEC of h-BN sheet using molecular dynamics and shows negative thermal expansion at low temperatures.

The contribution of in-plane and out-of-plane modes to the thermal expansion as functions of T for graphene are studied earlier [25] where the anharmonic couplings are determined from experiments. Negative thermal expansion (NTE) is crucial, especially since it is widely observed in almost all 2D materials. Moreover, NTE have physical origin at low temperature. At finite temperature, NTE is dominantly caused by the "vibrational elongation" effect owing to large out-of-plane fluctuation according to the calculation based on self-consistent phonon theory [26].

We have previously studied the phonon dispersion and various thermal properties like specific heat, entropy and free energy of pure graphene and how they varied with different B and N doping concentrations for graphene [27] where we also reported some estimates of LTEC for pure graphene. Since these estimates gave us negative values of thermal expansion, we now recalculate the estimates by carefully understanding the origin of negative expansion.

Therefore apart from study of thermal expansion, the primary goal of this paper has been to understand the reasons for negative thermal expansion. Since the calculation of thermal expansion involves summation over various phonon branches, therefore it is necessary to focus on individual phonon branches. Due to the 2-D structure of graphene with 2 atoms per unit cell, the 4 external phonon branches are usual with 2 acoustic and 2 optic ones. We calculate their contribution to thermal expansion separately. The other two branches in the transverse direction identified as ZA (transverse acoustic) and ZO (transverse optic) originate due to quasi-2D behavior and are very special modes of vibration for 2-D structures. Their restoring forces are derived from in plane modes and these behave very differently. Realizing this we have focused on their contribution separately and found that most important of this is transverse branch. We have extended the LTEC study of graphene by varying the boron and nitrogen dopant concentrations as well as of h-BN sheet.

The present paper is structured as follows. After introduction in section I where graphene and h-BN and their various properties are presented, a brief theory and computational details are summarized in Sec. II. Section III give results and discussion which includes lattice thermo-dynamical properties such as thermal expansion, mode Grüneisen parameters and surface bulk modulus as obtained from the variation of vibrational free energy with volume for pure as well as B and N doped graphene and h-BN sheet. Sec. IV contains summary and conclusions.

## 2. THEORY AND COMPUTATIONAL DETAILS

### 2.1) Theory

To calculate the phonon frequencies we have made use of VASP (Vienna *ab- initio* Simulation Package) [28-29] code based on density functional perturbation theory (DFPT) in interface with phonopy [30] code. Under quasi harmonic approximation, the implicit shift in the phonon frequencies arises due to the thermal expansion and is obtained from the volume dependence of the phonon frequencies, termed as Grüneisen parameter which is defined [31,32] as

$$\gamma_{\mathbf{q}j} = -\frac{\partial \ln \omega}{\partial \ln V} = -\frac{V_0}{\omega_{\mathbf{q}j}} \frac{\partial \omega_{\mathbf{q}j}}{\partial V} \qquad (1)$$

Where $V_0$ is equilibrium volume and $\omega_{\mathbf{q}j}$'s the harmonic phonon frequencies.

The mode Grüneisen parameters are the key ingredient in thermal expansion mechanisms. These parameters are generally positive but for some modes and for specific q values, they are found to be negative in a narrow temperature range. They are highly negative for lowest acoustic mode in case of graphene giving a negative thermal expansion at low temperatures. This may be due to that fact that only lower acoustic modes can be excited at low temperatures.

In absence of external pressure, the equilibrium structure of a crystal structure at any temperature T can be obtained by minimizing the free energy F=U-TS with respect to its degrees of freedom. Under harmonic approximation, free energy F is the sum of the ground state energy and the vibrational free energy which is the partition function of all independent harmonic oscillators. To obtain the lowest order contribution to thermal expansion, the quasi-harmonic free energy is written as [31]

$$F_{qh} = \phi(V) + k_B T \sum_{\mathbf{q},j} \ln\left[2\sinh\left(\frac{\hbar\omega_{\mathbf{q}j}}{2k_B T}\right)\right] \quad (2)$$

Where $\phi(V)$ is the static lattice contribution which can be written [28] as

$$\phi(V) = \phi(V_0) + \frac{1}{2}\varepsilon^2 V_0^2 \left(\frac{\partial^2 \phi}{\partial V^2}\right) = \phi(V_0) + \frac{1}{2}\varepsilon^2 V_0 B \quad (3)$$

Where the bulk modulus B is

$$B = V_0 \left(\frac{\partial^2 \phi}{\partial V^2}\right) \quad (4)$$

In case of 2D structure, we define surface bulk modulus as

$$B_S = S\left(\frac{\partial^2 \phi}{\partial S^2}\right) = LB \quad (5)$$

Where S is the surface area and L is an arbitrary length which needs to be introduced in the otherwise 3-D standard software.

Using condition for minimum energy, expression for thermal expansion coefficient β [derived in 31] is

$$\beta = \frac{k}{V_0 B} \sum_{\mathbf{q}j} \gamma_{\mathbf{q}j} \left(\frac{\hbar\omega_{\mathbf{q}j}}{k_B T}\right)^2 \frac{e^{\hbar\omega_{\mathbf{q}j}/kT}}{(e^{\hbar\omega_{\mathbf{q}j}/kT} - 1)^2} \quad (6)$$

LTEC ($\alpha$) (for 2D materials) is defined as

$$\alpha(T) = \frac{1}{a(T)}\frac{da(T)}{d(T)} = \frac{1}{2V}\frac{\partial V}{\partial T} = \frac{1}{2}\beta \quad (7)$$

As is evident from equation (6), the thermal expansion involves summation over all branches and the mode dependent Grüneisen parameters. It has been found that mode dependent Grüneisen parameter is very sensitive to the choice of branch. Therefore we introduce $\beta_j$ for each branch to calculate branch dependent thermal expansion such that

$$\beta = \sum_j \beta_j \quad (8)$$

**2.2) Computational parameters**

We adopted the Perdew-Burke-Ernzerhof (PBE) [33] exchange correlation (XC) functional of generalized gradient approximation (GGA) in our calculations. The plane wave cut-off energy was set to 750eV. The 4x4 supercell (consisting of 32 atoms) has been used to simulate the isolated graphene sheet and sheets are separated by larger than 10Å along perpendicular direction to avoid interlayer interactions.

The Monkhorst-pack scheme is used for sampling Brillouin zone. The structure is fully relaxed with Gamma centered 7 x 7 x 1 k-mesh. The partial occupancies were treated using the tetrahedron methodology with Blöchl corrections [34]. For geometry optimizations, all the internal coordinates were relaxed until the Hellmann-Feynman forces were less than 0.005 Å.

We have made use of phonopy software in combination with VASP to produce phonon frequencies. To obtain the force constants (FCs) in phonopy, we used density functional perturbation theory to calculate the IFC's of the perturbed structure. The phonon frequencies and surface bulk modulus are used in calculating LTEC using equations (6) & (7).

## 3. RESULTS AND DISCUSSION

### 3.1) Pure Graphene

The study of pure graphene under quasiharmonic aaproximation gives an estimate of Grüneisen parameters, bulk modulus and thermal expansion coefficient. The phonon dispersion for pure and doped graphene has been obtained and compared with various theoretical and experimental results in our previous studies [27].

#### 3.1.1) Energy-volume Curve

Under quasi harmonic approximation, we study the free energy of the structure at different unit cell volumes and found the E-V curve which is shown in Figure 1. It is to be noted that the volume here is representative of unit cell area. This first derivative of the energy-volume curve gives the minimum unit cell volume of 63.25 Å$^3$. The lattice constant corresponding to unit cell of this volume of graphene sheet is 2.47 Å. The minimum free energy is -18.47 eV. This value of lattice constant for graphene is very close to 2.46 Å as reported in literature [13].

Thus bulk modulus using equation (4) is obtained from second derivative of the fitted curve to the calculated E-V data as shown in Fig. 1. Hence we obtained surface bulk modulus for pure graphene to be equal to 90.24 N/m. This is further used in calculation of thermal expansion.

#### 3.1.2) Mode dependent Grüneisen parameter

The Grüneisen parameter plays a crucial role in determining temperature dependence of thermodynamic properties of materials ranging from phonon frequency shift to thermal expansion of a material. These parameters are generally positive, since phonon frequencies decrease when the solid expands, although some negative mode Grüneisen parameters for low frequency acoustic modes have been reported and usually combine with positive ones from different modes in the end diluting the effectiveness of positive Grüneisen parameters. Interestingly, thermal expansion can not only be diluted by these but also turn positive expansion to a negative expansion. To analyze thermal contraction, we have shown in Figure 2, mode dependent Grüneisen parameters of graphene. These are obtained from an interpolation of the phonon frequencies with a quadratic or linear polynomial in the lattice constants and computed at the ground state geometry. In graphene some bands display large and negative Grüneisen parameters, the Grüneisen parameters for the lowest acoustic branch of graphene is very low as compared with other branches. Therefore, at low temperatures where most optical modes with positive Grüneisen parameters are still not excited the contribution from the negative Grüneisen parameters will be dominant and thermal expansion would be negative. The negative Grüneisen parameters correspond to the lowest transversal acoustic ZA modes. These negative Grüneisen parameters will correspond to negative thermal expansion in graphene.

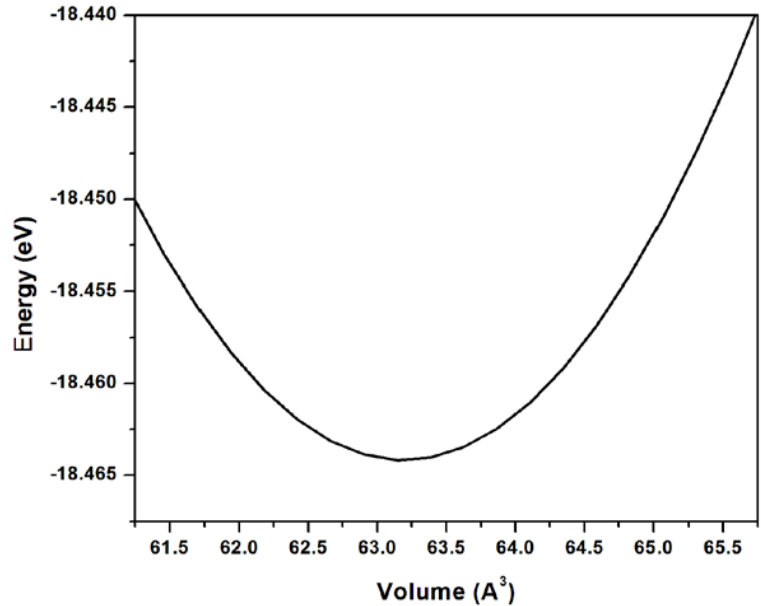

Figure 1 Energy-Volume Curve for pure graphene where volume is area multiplied by a constant height in z-direction

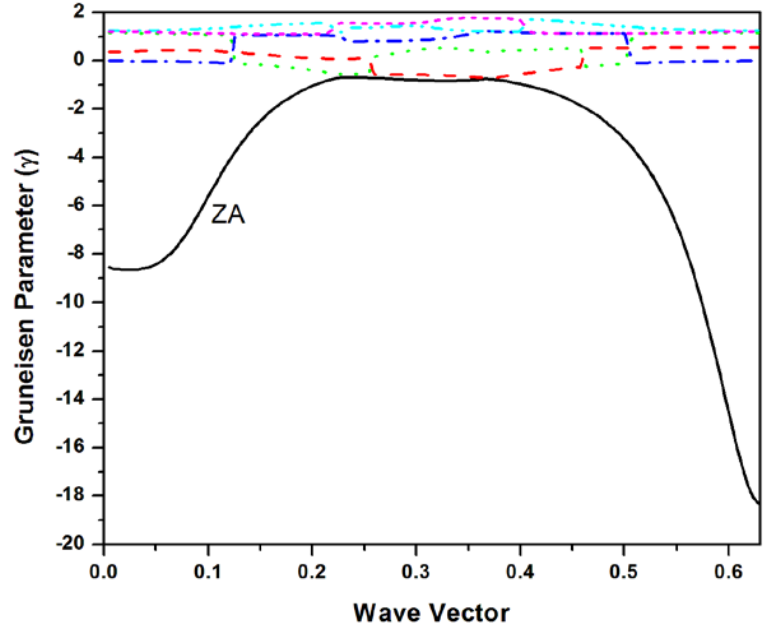

Figure 2 Grüneisen parameters for various phonon branches of graphene

### 3.1.3) Thermal Expansion Coefficient

The key feature of thermal properties of graphene is their extremely high thermal conductivity, which makes it a suitable material for heat control in high speed integrated electronic circuits. For such applications, knowledge of the thermal expansion coefficient (TEC) as a function of temperature is also important. Many authors have contributed to the calculation of TEC using various theoretical models and experimental techniques [10-15]. Bao *et al.* [14] experimentally measured the thermal coefficient in a very narrow temperature range of 300 – 400 K by monitoring the change in the sagging of a suspended graphene piece and found it to be negative only up to ~350 K with a room temperature value of -7 X $10^{-6}$ $K^{-1}$. Mounet *et al*. [10] estimated the LTEC of graphene using a first-principles calculation and predicted that graphene has negative LTEC in the whole temperature range under study (0-2500K).

There is a large discrepancy between theoretical and experimental data which may be caused by the inaccuracy of the experimental measurements or due to limitations of theoretical estimates. To get complete and accurate understanding of the behavior of LTEC near room temperature, which is important in designing graphene-based heat devices, we studied the branch dependent LTEC in detail.

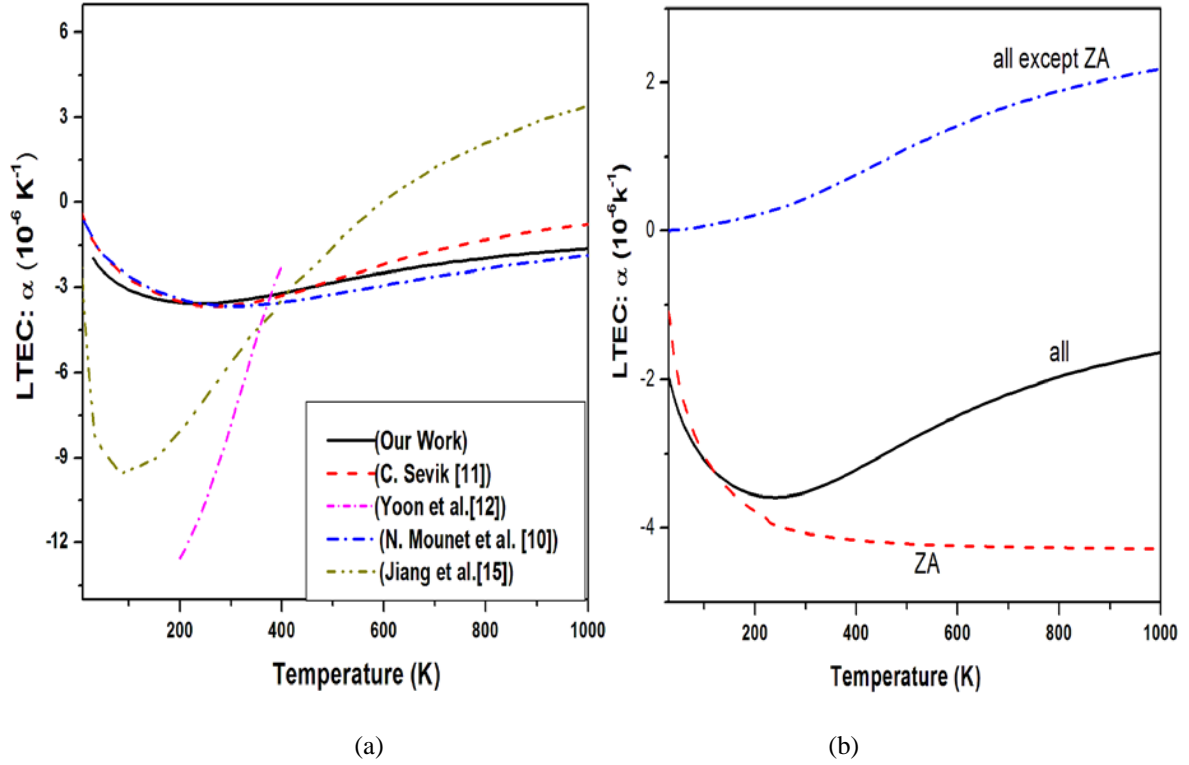

(a)                                             (b)

Figure 3(a) Temperature dependence of theoretical linear thermal expansion coefficient α of single layer graphene compared with experimental [12] (magenta short dash dot line) and theoretical measurements [10] ( blue dash dot line), [11] (red dash line) and [15] (dark yellow dash dot dot line). (b) Branch dependent thermal expansion

DFPT calculations of pure graphene for LTEC calculations show a negative behavior in whole range of temperature under study with a RT value of (-3.51X10$^{-6}$). In this calculation we have made use of surface bulk modulus calculated earlier using E-V curve. Our results are in good agreement with theoretical studies [10, 11] as shown in Figure 3(a) but there is still a discrepancy as compared to experimental results. To analyze it further, we present partial contributions to LTEC arising from each branch separately as the phonon branch ZA in particular has large negative Grüniesen parameters and must be visualized separately. Figure 3(b) shows the branch dependent LTEC which clearly chows that thermal expansion of graphene is negative only due to presence of large negative Grüneisen parameters corresponding to ZA branch. Here we have made use of equation (8) to calculate branch dependent thermal expansion. The major contribution to negative thermal expansion in graphene is contributed to ZA branch. The contribution from all other branches except ZA is almost similar which are clubbed together in a single branch and are all positive. The overall thermal expansion is negative due to dominant effect of ZA branch compared to other branches.

**3.2) Doped grapheme**

Since the B (N) doped structures are stable upto 25% doping [27], we have done the quasiharmonic study for two cases only i.e. 12.5% and 25% The atomic configurations corresponding to doped sheet is shown in Figure 4.

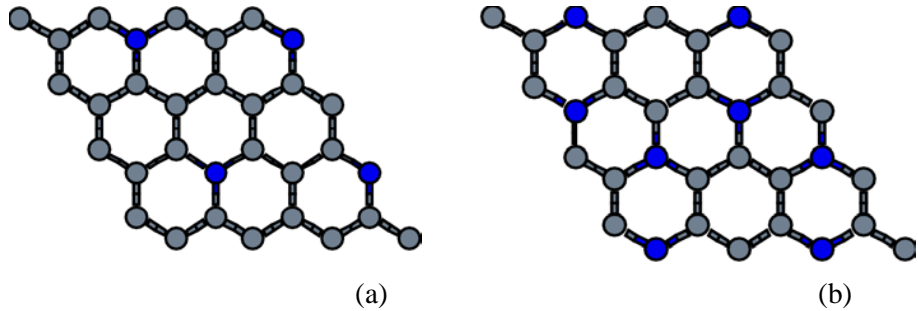

(a)                  (b)

Figure 4 Schematic diagrams of (a) 12.5% and (b) 25% doping concentrations

We have also considered the minimum doping configuration. The minimum doping can be one atom in a sheet of 32 atoms i.e. 3.125% in this case. But the configuration with low doping i.e. 1/2/3 atoms in a sheet is also thermodynamically unstable where almost all of the ZA mode frequencies are imaginary for boron doping and half of ZA frequencies are imaginary for nitrogen doping, as doping of B atoms introduces more lattice mismatch compared to N atoms doping which makes the structure thermodynamically more unstable. Thus considering one, two or three atoms doping in a hexagonal ring gives the thermodynamically unstable structures. We finally studied thermal expansion in only above two thermodynamically stable structures as already reported in our previous studies [27].

### 3.2.1) N-Doping

Thermal expansion coefficient has been calculated for the doped structure for different configurations and compared with that of pure graphene as shown in Figure 5. There is lowering of LTEC value with increasing doping concentration with a RT value of $-8.19 \times 10^{-6}$ for 12.5% doping and $-9.66 \times 10^{-6}$ for 25% doping. The large increase in negative value of thermal expansion with doping is primarily concerned with the ZA branch (similar to that shown earlier for pure graphene in figure 3(b)) which has more negative Grüneisen parameters near gamma point as low as -150 and -200 for 12.5% and 25% N doped sheets respectively. The surface bulk modulus values of doped sheets are also depicted using E-V curve to be 106.32N/m and 97.2 N/m for 12.5% and 25% N doped sheets respectively.

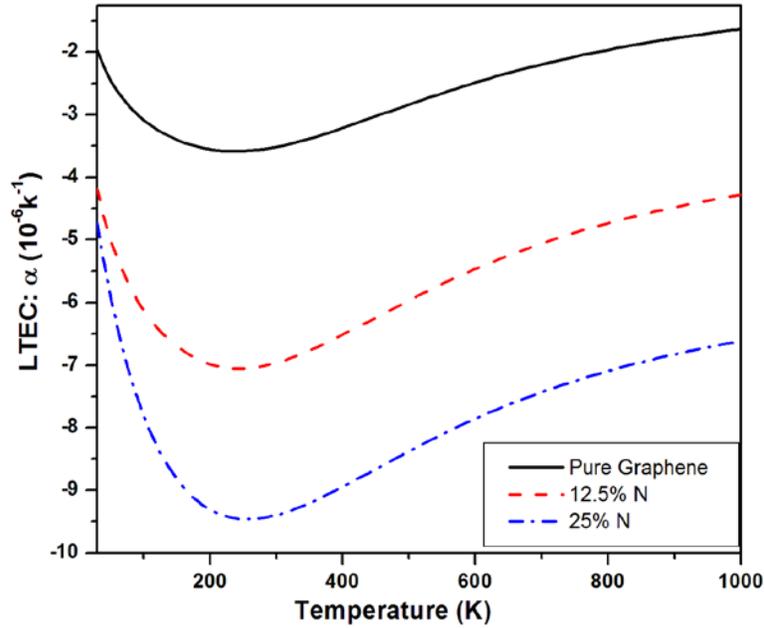

Figure 5 LTEC of 12.5% and 25% N-doped sheet compared with pure graphene case

### 3.2.2) B-Doping

Thermal expansion coefficient for the B doped structure for different doping configurations is shown in Figure 6. There is lowering of thermal expansion value with doping with a RT value of $-9.17 \times 10^{-6}$ for 12.5% doping and $-11.16 \times 10^{-6}$ for 25% doping. The increase in negative value of thermal expansion with B doping is comparatively larger than N doping although the trend is similar in both cases. The contribution of ZA branch in thermal expansion is to make it negative as the Grüneisen parameters for ZA branch for the B doped structures are as low as -105 and -150 for 12.5% and 25% respectively. However the increase in negative value of LTEC for the case of B doped sheet as compared to N doped sheet is much larger due to decrease in its surface bulk modulus value. The surface bulk modulus values of B doped sheets are depicted to be 92.5 N/m and 73.5N/m for 12.5% and 25% doping configurations respectively.

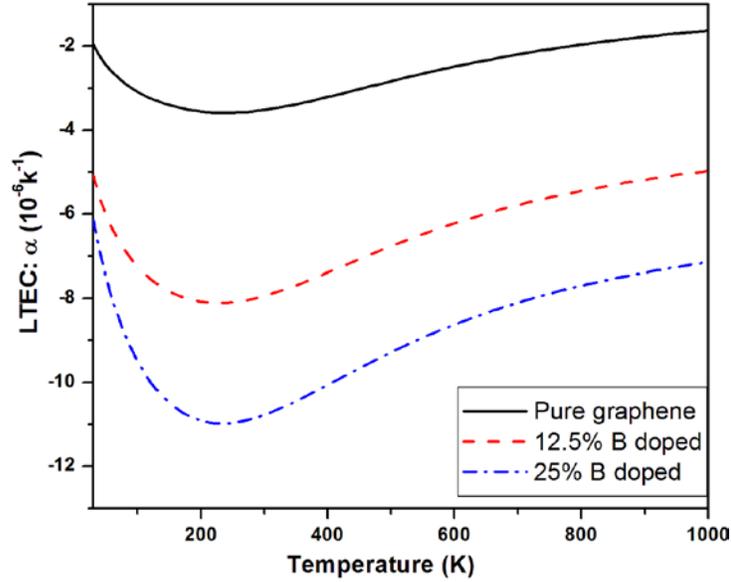

Figure 6  LTEC of 12.5% and 25% B-doped sheet compared with pure graphene case

### 3.3) h-BN sheet

To understand thermodynamic properties of h-BN sheet, the first step is to inspect lattice vibrational modes (phonons).

### 3.3.1) Phonon Dispersion

Similar to phonon dispersion of graphene calculated earlier [27], h-BN sheet also have 6 phonon branches (three acoustic (A) and 3N-3 optical (O) phonon modes).

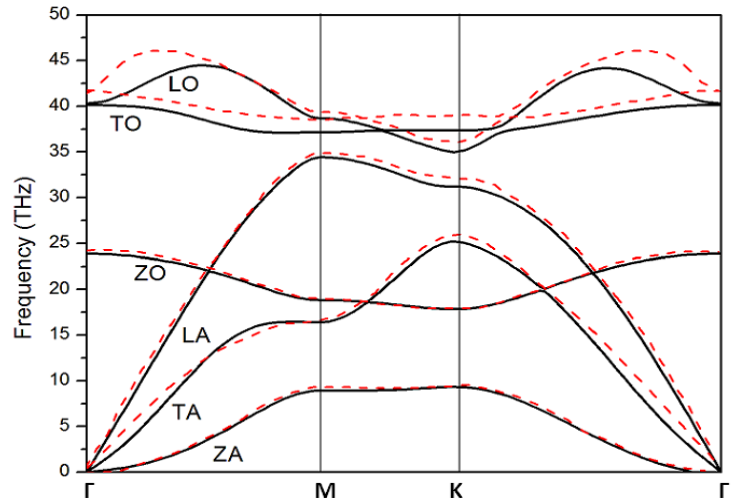

Figure 7 Phonon Dispersion curve for h-BN (solid lines) as compared to that in Ref [32] (red dashed lines). LA (LO), TA (TO) and ZA (ZO) are longitudinal, transversal and out-of-plane acoustical (optical) branches respectively.

The phonon frequencies are obtained under harmonic approximation which gives phonon dispersion curve of h-BN sheet as shown in Figure 7. The curves have been obtained by simulating the h-BN sheet for directions between the high-symmetry Γ point and a large number of points on the line. We have made use of density functional perturbation theory with displacement along all the three directions which give rise to 6 phonon branches in dispersion curve... The phonon dispersion curve is in good agreement with reference [35] where Wirtz *et al.* have studied the phonon dispersion curve using molecular dynamics. Our phonon dispersion curve is also fairly in agreement with experimental studies [36, 37] done using inelastic x-ray scattering and Raman analysis methods.

### 3.3.2) Energy volume curve

Similar to the pure graphene, the E-V curve obtained for h-BN sheet is shown in figure 8. The volume (area of sheet multiplied by a constant height along z-axis) which minimizes the curve is 55.26 Å$^3$.

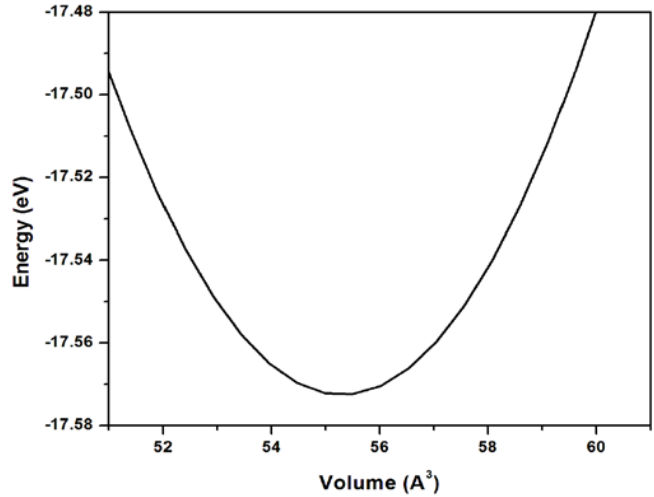

Figure 8 Energy-Volume Curve for h-BN sheet

The lattice constant corresponding to minimum unit cell volume is 2.515 Å. The surface bulk modulus found to be 74.6N/m for h-BN sheet which is much smaller than surface bulk modulus of pure graphene.

### 3.3.3) Grüneisen parameter

The mode dependent Grüneisen parameter for h-BN sheet has been obtained for various phonon branches corresponding to different q-values and shown in Figure 9(a). The Grüneisen parameter corresponding to lowest transverse acoustic mode is negative in the whole region with lowest values approaching as low as – 230 larger than doped and pure graphene case as shown in Figure 9(b). There is a gap between ZA mode Grüneisen parameters and other acoustic and optical mode Grüneisen parameters contrary to the case of graphene as shown in Figure 2 where there is no gap. This observation is similar to the observation of difference in phonon dispersion curve of graphene and h-BN.

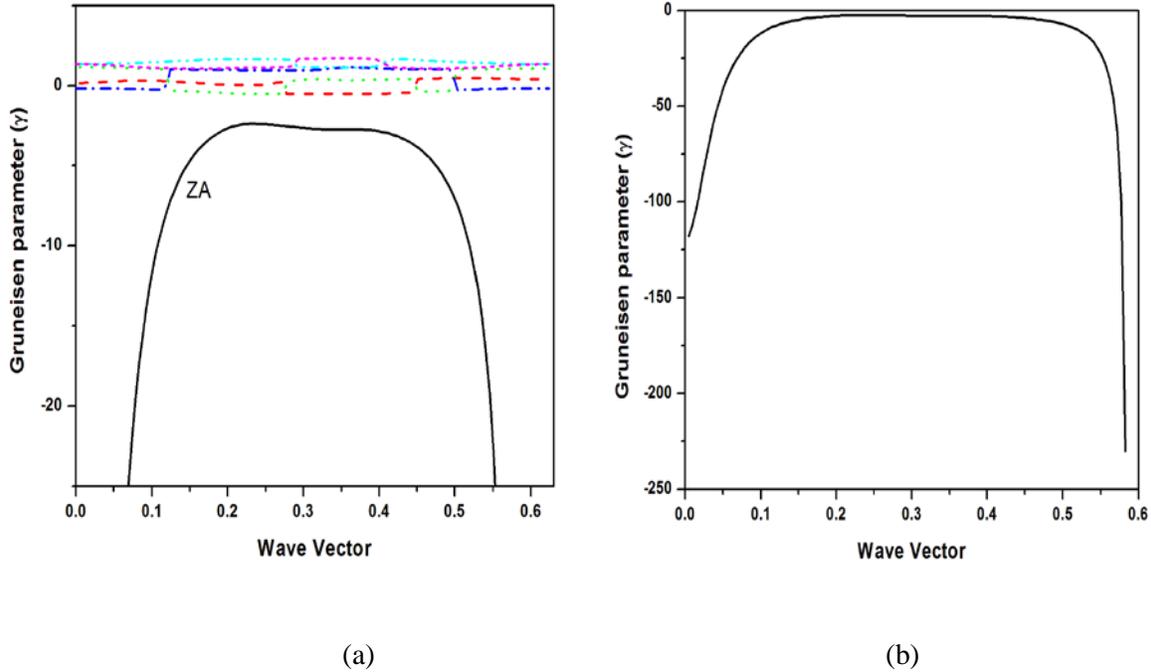

(a) (b)

Figure 9(a) Mode dependent Grüneisen parameters, (b) Grüneisen parameter of ZA mode

**3.3.4) Thermal Expansion Coefficient**

For calculating LTEC of single layer h-BN sheet, several authors have used various methods [11, 24, 38-40] using various experimental techniques [38-40] and theoretical models [11, 24]. Sevik [11] theoretically estimated the LTEC in the temperature range of 0-1500 K by doing calculations using DFPT as implemented in VASP software. Thomas *et al.* [24] have investigated LTEC in a wide temperature range by carrying out atomistic (molecular dynamics) simulations using a tuned Tersoff-type inter-atomic empirical potential. The results of various experimental techniques for bulk h-BN crystals [35-37] are in the temperature range 0-400K and lower than the theoretical single sheet estimates given by Sevik [11].

We obtained the behavior of LTEC for h-BN sheet using DFPT under QHA as shown in Figure 10(a). Although the room temperature value of LTEC is much larger than that determined by Sevik [11] of the order of -12.86x10-6 but the behavior is in conformity with B and N doped structures. The negative thermal expansion can be explained by the existence of more negative Grüneisen parameters corresponding to the lowest transverse acoustic (ZA) and lower surface bulk modulus compared to graphene. The experimentally measured values in the temperature range 0–400 K are for bulk h-BN crystals which are lower than estimates for single h-BN sheet. The prominent difference in estimates arises due to the strong anharmonic character of h-BN structure, which is not taken into account in QHA calculations. Alternately, the LTEC values predicted by Thomas et al. [24] based on fully atomistic simulations gives a lower estimate compared with Sevik [11]. Figure 10(b) shows branch dependent LTEC which depicts the contribution of ZA branch is of utmost importance to total thermal expansion as compared to other branches.

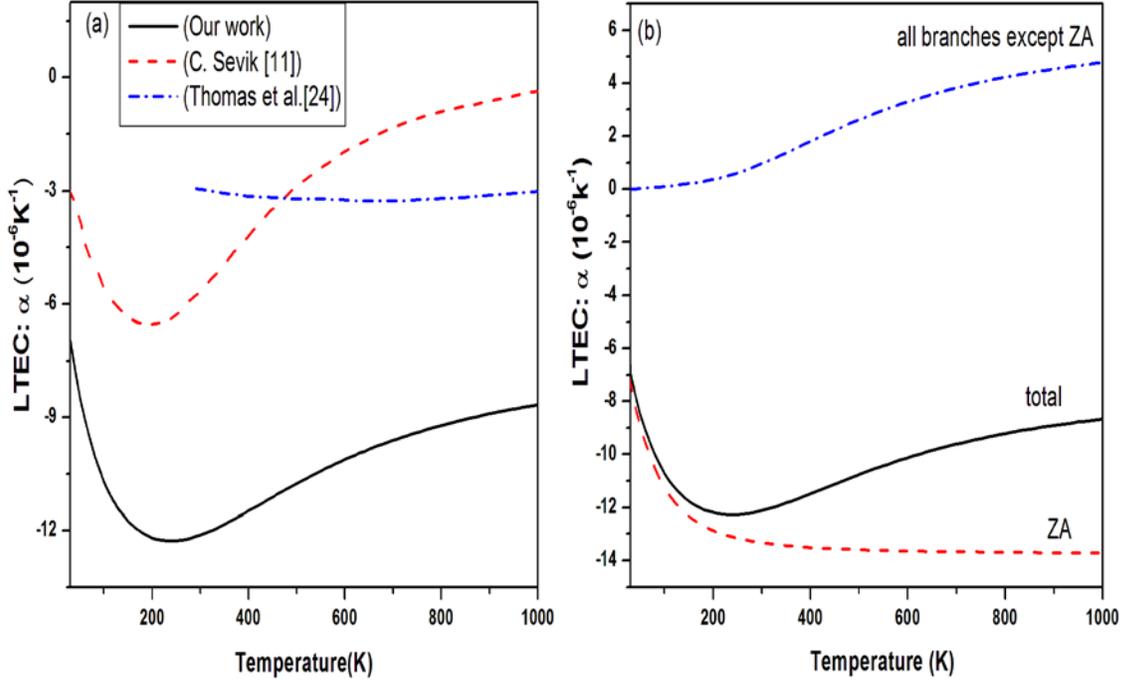

Figure 10(a) Temperature dependence of LTEC for single layer *h*-BN black solid line. Our results are compared with various theoretical measurement (Ref [11] red dashed line and Ref [24] blue dash dot line). (b) Branch dependent thermal expansion

Figure 11 shows a comparison of LTEC for pure and doped graphene compared with h-BN sheet. The graph trend shows an increase in negative thermal expansion with increase in N and B doping and further increases in h-BN sheet. We also observe that B doping makes the LTEC value more negative as compared to same doping percentage of N atom. Further we notice that our estimates are overestimating the LTEC values for h-BN sheet when compared to [11] by a factor close to 2. However, our study of systematic doping by increasing B and N concentrations follows the pattern of LTEC as shown in figure 11 quite well where we present results for doping of graphene with individual B and N atoms going from 12.5% to 25% and finally 50% B and N doping together creating h-BN sheet.

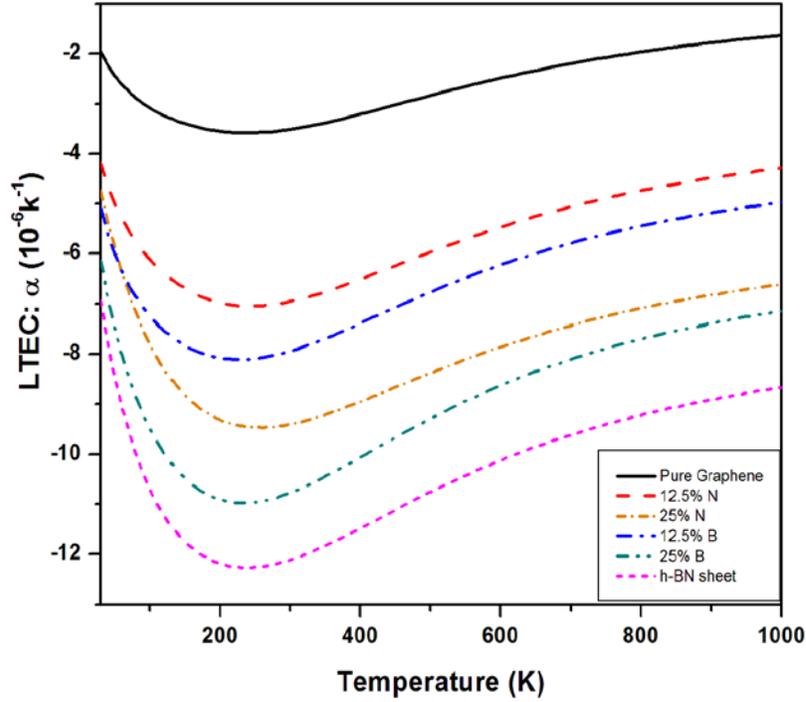

Figure 11 Temperature dependence of LTEC for pure graphene compared with B and N doped graphene and single layer *h*-BN sheet

## 4) SUMMARY AND CONCLUSIONS

We have presented a detailed first-principles study at the GGA-PBE level using the quasi-harmonic approximation to derive the finite temperature behavior of thermal expansion in pure graphene. The thermal expansion coefficient is negative in the whole temperature region under study and is in qualitatively good agreement with available theoretical data although it deviates from experimental studies. We have also presented study of *h*-BN phonon dispersion and derived the linear thermal expansion coefficient. The phonon frequencies and phonon dispersions are in good agreement with various theoretical and experimental results.

A comparison of LTEC for pure, B/N doped graphene and h-BN sheet has been made. There is similar trend of increase in negative Grüneisen parameters with B/N doping which continues for h-BN sheet as well when compared with pure graphene. There are quantitative differences between various theoretical estimates of thermal expansion as well as with experimental data. The convergence of various results will be more in focus when experimental estimates are also presented more consistently. But the fact that there is a large negative LTEC for all 2-D systems and that ZA which is transverse acoustic mode is the "culprit" in dictating drastic unusual behavior is a clear message of this paper. Therefore on the basis of the results gathered here and earlier, one can safely conclude that 2-D structures do have tendencies to show negative thermal expansion. The study of LTEC for B/N doped graphene is entirely new and the doped structures can be designed to get required thermal expansion for specific applications or combined with different materials to compensate the positive expansion to designable zero or either kind of thermal expansion materials.

## 5) ACKNOWLEDGEMENTS

We express our gratitude to VASP team for providing the code, the phonopy team for the code, the HPC facilities at IUAC (New Delhi) and the departmental computing facilities at Department of Physics, Panjab University, Chandigarh. VKJ gratefully acknowledges support from Alexander von Humboldt for support as a Guest Professor at University of Würzburg.